\documentclass[12pt,epsf]{article}
\usepackage{graphicx,amsmath,amssymb}
\begin{document}
%%%%%%%%%%%%%%%%%%%%%%%%%%%%%%%%%%%%%%%%%%%

\def\a{\alpha}
\def\b{\beta}
\def\c{\varepsilon}
\def\d{\delta}
\def\e{\epsilon}
\def\f{\phi}
\def\g{\gamma}
\def\h{\theta}
\def\k{\kappa}
\def\l{\lambda}
\def\m{\mu}
\def\n{\nu}
\def\p{\psi}
\def\q{\partial}
\def\r{\rho}
\def\s{\sigma}
\def\t{\tau}
\def\u{\upsilon}
\def\v{\varphi}
\def\w{\omega}
\def\x{\xi}
\def\y{\eta}
\def\z{\zeta}
\def\D{\Delta}
\def\G{\Gamma}
\def\H{\Theta}
\def\L{\Lambda}
\def\F{\Phi}
\def\P{\Psi}
\def\S{\Sigma}

\def\o{\over}
\def\beq{\begin{eqnarray}}
\def\eeq{\end{eqnarray}}
\newcommand{\gsim}{ \mathop{}_{\textstyle \sim}^{\textstyle >} }
\newcommand{\lsim}{ \mathop{}_{\textstyle \sim}^{\textstyle <} }
\newcommand{\vev}[1]{ \left\langle {#1} \right\rangle }
\newcommand{\bra}[1]{ \langle {#1} | }
\newcommand{\ket}[1]{ | {#1} \rangle }
\newcommand{\EV}{ {\rm eV} }
\newcommand{\KEV}{ {\rm keV} }
\newcommand{\MEV}{ {\rm MeV} }
\newcommand{\GEV}{ {\rm GeV} }
\newcommand{\TEV}{ {\rm TeV} }
\def\diag{\mathop{\rm diag}\nolimits}
\def\Spin{\mathop{\rm Spin}}
\def\SO{\mathop{\rm SO}}
\def\O{\mathop{\rm O}}
\def\SU{\mathop{\rm SU}}
\def\U{\mathop{\rm U}}
\def\Sp{\mathop{\rm Sp}}
\def\SL{\mathop{\rm SL}}
\def\tr{\mathop{\rm tr}}

\def\IJMP{Int.~J.~Mod.~Phys. }
\def\MPL{Mod.~Phys.~Lett. }
\def\NP{Nucl.~Phys. }
\def\PL{Phys.~Lett. }
\def\PR{Phys.~Rev. }
\def\PRL{Phys.~Rev.~Lett. }
\def\PTP{Prog.~Theor.~Phys. }
\def\ZP{Z.~Phys. }

%%%%%%% added by Fumi %%%%%%%%%%
% FROM HERE
%\newcommand{\beq}{\begin{equation}}   
%\newcommand{\eeq}{\end{equation}}
\newcommand{\bea}{\begin{eqnarray}}   
\newcommand{\eea}{\end{eqnarray}}
\newcommand{\bear}{\begin{array}}  
\newcommand {\eear}{\end{array}}
\newcommand{\bef}{\begin{figure}}  
\newcommand {\eef}{\end{figure}}
\newcommand{\bec}{\begin{center}}  
\newcommand {\eec}{\end{center}}
\newcommand{\non}{\nonumber}  
\newcommand {\eqn}[1]{\beq {#1}\eeq}
\newcommand{\la}{\left\langle}  
\newcommand{\ra}{\right\rangle}
\newcommand{\ds}{\displaystyle}
\def\SEC#1{Sec.~\ref{#1}}
\def\FIG#1{Fig.~\ref{#1}}
\def\EQ#1{Eq.~(\ref{#1})}
\def\EQS#1{Eqs.~(\ref{#1})}
\def\GEV#1{10^{#1}{\rm\,GeV}}
\def\MEV#1{10^{#1}{\rm\,MeV}}
\def\KEV#1{10^{#1}{\rm\,keV}}
\def\lrf#1#2{ \left(\frac{#1}{#2}\right)}
\def\lrfp#1#2#3{ \left(\frac{#1}{#2} \right)^{#3}}
% UNTIL HERE

%%%%%%%%%%%%%%%%%%%%%%%%%%%%%%%%%%%%%%%%%%%%%%%%%%%%%%%%%%%%%%%%%%%%

\baselineskip 0.7cm

\begin{titlepage}

\begin{flushright}
UT-13-39\\
TU-950\\
IPMU13-0222\\
\end{flushright}

\vskip 1.35cm
\begin{center}
{\large \bf 
Chaotic Inflation with Right-handed Sneutrinos after Planck
}
\vskip 1.2cm
Kazunori Nakayama$^{a,c}$,
Fuminobu Takahashi$^{b,c}$
and 
Tsutomu T. Yanagida$^{c}$

\vskip 0.4cm
{\it $^a$Department of Physics, University of Tokyo, Tokyo 113-0033, Japan}\\
{\it $^b$Department of Physics, Tohoku University, Sendai 980-8578, Japan}\\
{\it $^c$Kavli Institute for the Physics and Mathematics of the Universe (WPI), TODIAS, University of Tokyo, Kashiwa 277-8583, Japan}

\vskip 1.5cm

\abstract{
We propose a chaotic inflation model in which the lightest right-handed sneutrino serves as the inflaton 
and the predicted values of the spectral index and tensor-to-scalar ratio are
consistent with the Planck data. Interestingly, the observed magnitude
of primordial density perturbations is naturally explained by the inflaton mass of order $\GEV{13}$,
which is close to the right-handed neutrino mass scale suggested by the seesaw mechanism and 
the neutrino oscillation experiments. We find that the agreement of the two scales becomes even better in the neutrino mass
anarchy. 
We show that the inflation model can be embedded into supergravity  and discuss
 thermal history of the Universe after inflation such as non-thermal leptogenesis by the 
 right-handed sneutrino decays and the modulus dynamics. 
}
\end{center}
\end{titlepage}

\setcounter{page}{2}

%%%%%%%%%%%%%%%%%%%%%%%%%%%%%%%%%%%%%
\section{Introduction}
%%%%%%%%%%%%%%%%%%%%%%%%%%%%%%%%%%%%%

The Planck results~\cite{Ade:2013rta} confirmed the vanilla $\Lambda$CDM model with six
cosmological parameters based on almost scale-invariant, adiabatic and Gaussian primordial 
density perturbations. This strongly suggests that our Universe experienced 
the inflationary epoch described by a simple (effectively) single-field inflation~\cite{Guth:1980zm,Linde:1981mu}.

The primordial density perturbations are parametrized by the spectral index $n_s$
and the tensor-to-scalar ratio $r$, and they are tightly constrained by 
the Planck data combined with other CMB and cosmological observations.
Roughly speaking, $n_s$ and $r$ are sensitive to the shape and magnitude of 
the inflaton potential, respectively. It is known that $r$ is related to the
field excursion of the inflaton, and the on-going and planned CMB
observations will be able to probe $r \gtrsim 10^{-3}$, for which
the inflaton field excursion exceeds the Planck scale. One of the large-field
inflation is the chaotic inflation~\cite{Linde:1983gd}. Intriguingly, the chaotic inflation based on the
monomial potential is outside the $1\sigma$ allowed region, and in particular,
the quadratic chaotic inflation is near the boundary of the $2\sigma$ allowed region.
Interestingly, it was recently pointed out in Refs.~\cite{Croon:2013ana,Nakayama:2013jka,
Ellis:2013xoa,Kallosh:2013pby} (see Refs.~\cite{Destri:2007pv,Linde:1983fq,Kallosh:2007wm} for early attempts)
that the predicted values of $(n_s, r)$ lie inside the region allowed by the Planck data, if
the quadratic inflaton potential is slightly modified at large field values.

While scalar fields are ubiquitous in theories beyond the standard model (SM) such as supersymmetry 
(SUSY) or string theory,  the identity of the inflaton and its couplings to the SM sector are unknown.
Here we consider a model in which one of the right-handed sneutrinos  plays a role of the inflaton~\cite{Murayama:1992ua,Murayama:1993xu,Ellis:2003sq,Antusch:2009ty},
extending the original model by introducing a slight modification to the quadratic potential at large field values. 
The observed magnitude of the primordial density perturbations can be
naturally explained by the sneutrino mass of $\GEV{13}$, which is close to the right-handed
neutrino mass scale suggested by the seesaw mechanism~\cite{seesaw} 
and the neutrino oscillation experiments. We will show that the agreement will be even better in the neutrino mass anarchy 
hypothesis in which  the right-handed neutrino mass matrix is given by a random matrix~\cite{Hall:1999sn,Haba:2000be}. 
This model has an advantage over singlet inflation models, in that the inflaton has couplings with the
leptons and Higgs fields, which enable the successful reheating. 
Moreover the baryon asymmetry generation through leptogenesis~\cite{Fukugita:1986hr} naturally takes place.

%%%%%%%%%%%%%%%%%%%%%%%%%%%%%%%%%%%%%
\section{Chaotic inflation with right-handed sneutrinos}
%%%%%%%%%%%%%%%%%%%%%%%%%%%%%%%%%%%%%
\subsection{A model in global SUSY}
%%%%%%%%%%%%%%%%%%%%%%%%%%%%%%%%%%%%%
Let us first consider a chaotic inflation model with right-handed sneutrinos 
in a global SUSY framework. We will see shortly that it is possible to embed the
model into supergravity without significant modifications. 

We start with the following superpotential;
\beq
\label{ourW}
W\;=\; \frac{1}{2} M_{ij} N_i N_j + \frac{1}{4} \lambda_{ijkl} N_i N_j N_k N_l + \cdots,
\eeq
where $N_i$ denotes a chiral superfields for the $i$-th right-handed neutrino, $M_{ij}$ and $\lambda_{ijkl}$ represent the mass and quartic coupling of
the right-handed neutrinos, and  the flavor indices are $i,j  =1, 2, 3$. For the moment we assume a minimal K\"ahler potential
for $N_i$. Here and in what follows we adopt the Planck units where the reduced Planck mass
$M_p \simeq 2.4 \times \GEV{18}$ is set to be unity, unless explicitly shown otherwise for convenience.

In Ref.~\cite{Croon:2013ana} the inflation model with the  superpotential 
\[W = \frac{\mu}{2} N^2 - \frac{\lambda}{3} N^3\]
has been proposed under the name
of Wess-Zumino inflation, in which an R-parity is explicitly broken. One of the advantages of our model (\ref{ourW}) is that
the R-parity is preserved.

In order to estimate the size of the interactions, let us express $M_{ij}$ and $\lambda_{ijkl}$ as
\bea
M_{ij} &=& x_{ij} \Phi,\\
\lambda_{ijkl} &=& y_{ijkl} \Phi^2,
\eea
where $x_{ij}$ and $y_{ijkl}$ are numerical coefficients of order unity, $\Phi$ is a spurion field with ${\rm B-L}$ charge $+2$, and its expectation value
represents the magnitude of the ${\rm B-L}$ breaking.  To be concrete, we set $\Phi$ to be ${\cal O}(10^{-4})$ as suggested 
by the seesaw mechanism~\cite{seesaw} and the neutrino oscillation experiments.

 The flavor structure is represented by
$x_{ij}$ and $y_{ijkl}$, and we presume that they are complex-valued  random
matrices whose elements are of order unity, based on the
neutrino mass anarchy hypothesis~\cite{Hall:1999sn,Haba:2000be,Jeong:2012zj}. 
 It is known that the observed large mixing angles for neutrinos 
and the mild hierarchy for the mass squared differences can be nicely explained in the
neutrino mass anarchy hypothesis.

%%%%%%%%%%%%%%%%
\begin{figure}[t]
\begin{center}
\includegraphics[scale=0.8]{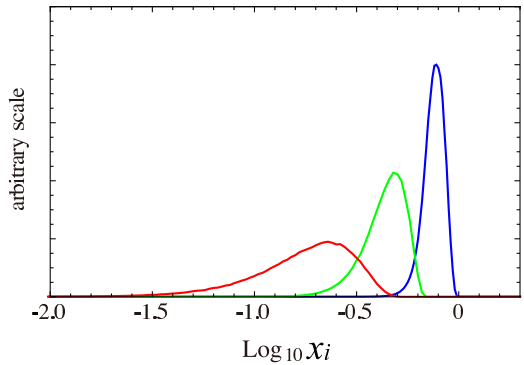}
\caption { 
	Probability distribution of the eigenvalues $(x_1,x_2,x_3)$ of the complex-valued symmetric random matrix $x_{ij}$,
	satisfying ${\rm Tr}[x^\dag x] \leq 1$ and $x_1 \leq x_2 \leq x_3$. The smallest eigenvalue $x_1$ 
	ranges from $10^{-1.5} (\sim 0.03)$
	to $10^{-0.3} (\sim 0.5)$.
	}
\label{fig:dist}
\end{center}
\end{figure}
%%%%%%%%%%%%%%%%

Let us go to the mass eigenstate basis, $\{\hat{N}_1, {\hat N}_2, {\hat N}_3\}$, with mass eigenvalues $M_1 \leq
M_2 \leq M_3$.
We identify the lightest right-handed sneutrino with the inflaton. Fig.~\ref{fig:dist} shows the probability 
distribution of the eigenvalues of the complex-valued symmetric random matrix $x_{ij}$.
As one can see the figure that the smallest eigenvalue typically ranges from $0.03$ to $0.5$. 
On the other hand, the flavor structure of the quartic couplings $\lambda_{ijkl}$ are independent of the
mass eigenstates, and so, we expect that $|y_{1111}| \sim 1$ in this basis.
Thus, the superpotential for the inflaton $\phi \equiv {\hat N}_1$ is given by
\beq
W \;=\; \frac{1}{2} M \phi^2 + \frac{1}{4} \lambda \phi^4,
\label{W}
\eeq
with
\bea
\label{M}
M &\equiv & M_1 \sim 0.1 \Phi \sim 10^{-5}, \\
\lambda & \equiv & \lambda_{1111} \sim \Phi^2 \sim 10^{-8},
\label{lam}
\eea
where we have dropped higher order terms, and we set $M$ and $\lambda$ real and positive for simplicity. 
 We also assume that,  during inflation,  the heavier two mass eigenstates ${\hat N}_2$ and ${\hat N}_3$ are
stabilized at SUSY minimum by their couplings with the inflaton ${\hat N}_1$.
The inflaton potential is given by
\beq
V(\phi) \;=\; \left|M \phi + \lambda \phi^3\right|^2 = M^2 |\phi|^2 + \lambda M |\phi|^2 \left(\phi^2 + \phi^{*2} \right) + \lambda^2 |\phi|^6
\eeq
Writing the inflaton field as $\phi = \varphi/\sqrt{2} e^{i \theta}$, the inflaton potential is minimized along $\cos 2\theta = 0$,
namely, $\theta = \pi/4$. The inflaton potential along the radial component is
\bea
V(\varphi) &=& \frac{1}{2} M^2 \varphi^2 - \frac{1}{2} \lambda M \varphi^4 + \frac{1}{8} \lambda^2 \varphi^6,\non\\
&=& \frac{1}{2} \left(M \varphi - \frac{1}{2} \lambda \varphi^3 \right)^2.
\label{V}
\eea
It is a shifted version of the symmetry breaking potential. 
The inflation is possible if it initially sits in the vicinity of the local maximum at $\varphi = \sqrt{2M/\lambda}$. 

\begin{figure}[t]
\begin{center}
\includegraphics[scale=1.5]{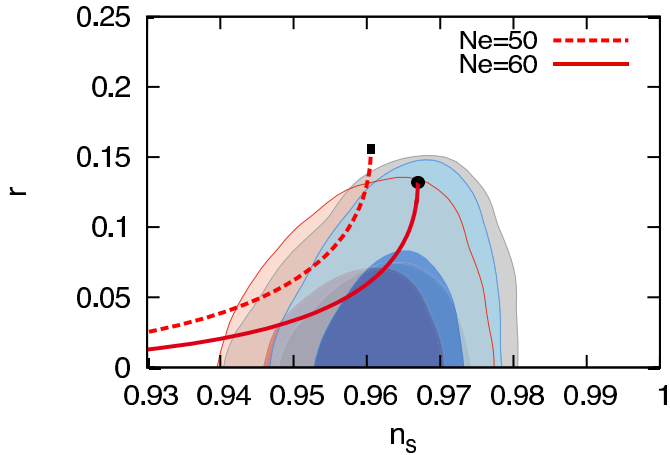}
\caption { 
	The prediction for $(n_s,r)$ is shown by the red lines for $N_e=50$ (dashed) and $N_e=60$ (solid). 
	The black points correspond to the case of chaotic inflation with quadratic potential.
	Together shown are the Planck constraint~\cite{Ade:2013rta}.
}
\label{fig:nsr}
\end{center}
\end{figure}

\begin{figure}[t]
\begin{center}
\includegraphics[scale=1.5]{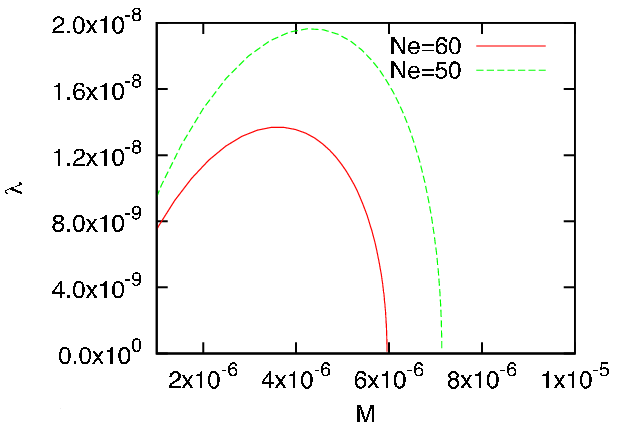}
\caption { 
	Parameters $M$ and $\lambda$ (in Planck unit) which reproduced the Planck normalization on the density perturbation 
	for $N_e=50$ (dashed) and $N_e=60$ (solid).
}
\label{fig:mlam}
\end{center}
\end{figure}

We have numerically solved the inflaton dynamics and calculated the predicted $n_s$ and $r$
as shown by the red lines in Fig.~\ref{fig:nsr} for the total e-folding number $N_e = 50$ and $60$.
The e-folding number depends on both the inflation scale and the thermal history after inflation.
If there is a late-time entropy production by e.g. modulus decay, the e-folding number becomes 
smaller. As we shall see shortly, there is a modulus when we embed the present model into supergravity. 
Then the e-folding number is given $N_e \simeq 54 - 55$, somewhere between the two lines.
The black points correspond to the case of chaotic inflation with quadratic potential.
One can see that, compared to the original quadratic chaotic inflation, the predicted $n_s$ and $r$ become smaller,
thanks to the higher order terms in the inflaton potential. 
We have imposed the Planck normalization of the primordial density perturbations,
and show how the parameters $M$ and $\lambda$ change in Fig.~\ref{fig:mlam}. 
Interestingly, the expected size of $M$ and $\lambda$
given by Eqs.~(\ref{M}) and (\ref{lam}) nicely match with the ranges favored by the Planck data.
Note that the neutrino mass anarchy improves the agreement between the seesaw scale and the inflaton mass.

%%%%%%%%%%%%%%%%
%%%%%%%%%%%%%%%%

%%%%%%%%%%%%%%%%
%%%%%%%%%%%%%%%%

%%%%%%%%%%%%%%%%%%%%%%%%%%%%%%%%%%%%%
\subsection{Embedding in supergravity}
%%%%%%%%%%%%%%%%%%%%%%%%%%%%%%%%%%%%%

We consider the following K\"ahler and super-potentials~\cite{Murayama:1993xu};
\bea
\label{Km}
K &=& \frac{3}{8} \ln \eta + \eta^2,\\
W &=&W(\phi_i)
\eeq
with $\eta \equiv z+ z^\dag + |\phi_i|^2$, where $z$ is a modulus field and $\phi_i$
denotes chiral superfields in the model. Later we will identify $\phi_i$ with the right-handed neutrinos. 
The coefficients in $K$ are chosen so that $\eta$ is stabilized
at $\eta = 3/4$ where the scalar potential vanishes~\cite{Murayama:1993xu}. In fact, 
a more general class of the K\"ahler and super-potentials can lead to successful chaotic inflation,
as we will show in Appendix. There are two important assumptions. One is that the K\"ahler potential
is written as a function of only $\eta$. The other is that the cosmological constant vanishes in the vacuum 
at the tree level.  So, if we allow another  up-lifting sector, successful chaotic inflation is possible
for an even broader class of  K\"ahler and super-potentials. To be concrete, however, we will focus
on the K\"ahler and super-potentials given above. 
%In the presence of another up-lifting sector, different numerical coefficients are possible.

The Lagrangian is given by
\bea
{\cal L} &=& \frac{16 \eta^2 -3}{32 \eta^2} \left( \partial_\mu \eta\, \partial^\mu \eta+I_\mu I^\mu \right)
+ \frac{16 \eta^2+3}{8 \eta} \partial_\mu \phi_i^* \partial^\mu \phi_i - V
\eea
with
\beq
V\;=\; \eta^\frac{3}{8} e^{\eta^2} \left(
\frac{8 \eta}{16 \eta^2+3} |W_i|^2+ \frac{(16\eta^2-9)^2}{8 (16\eta^2-3)} |W|^2
\right),
\eeq
where $W_i \equiv \partial W / \partial \phi_i$, and $I_\mu \equiv i \partial_\mu(z-z^*) + i (\phi_i^* \partial_\mu \phi_i
-\phi_i \partial_\mu \phi_i^*)$. 
 We have assumed that the right-handed neutrinos are gauge-singlet and there is no D-term potential.
It was shown in Ref.~\cite{Murayama:1993xu} that
the modulus $\eta$ is successfully stabilized during and after inflation, leading to the scalar
potential with the same form in the global SUSY. The effective potential for $\phi_i$ is given by
\beq
V =  \frac{1}{2}\lrfp{3}{4}{\frac{3}{8}} e^\frac{9}{16} \left|\partial_{\phi_i} W \right|^2.
\label{Vsugra}
\eeq
Then, assuming the superpotential of (\ref{W}), we obtain the inflaton potential (\ref{V})
after a trivial change of normalization of $M$ and $\lambda$ due to the numerical coefficient
appearing in \EQ{Vsugra}.
The predicted values of the spectral index and the tensor-to-scalar ratio are the same as in Fig.~\ref{fig:nsr}.

%%%%%%%%%%%%%%%%%%%%%%%%%%%%%%%%%%%%%
\section{Cosmology after inflation and phenomenological implications}
%%%%%%%%%%%%%%%%%%%%%%%%%%%%%%%%%%%%%

%%%%%%%%%%%%%%%%%%%%%%%%%%%%%%%%%%%%%
\subsection{Leptogenesis from inflaton decay}
%%%%%%%%%%%%%%%%%%%%%%%%%%%%%%%%%%%%%

The inflaton, i.e. the lightest right-handed sneutrino, decays into leptons and Higgs 
through the yukawa coupling
\begin{equation}
	W = h_{1\alpha} {\hat N}_1 L_\alpha H_u,
\end{equation}
where $L_\alpha$ and $H_u$ are chiral superfields for the
lepton doublet and up-type Higgs. The reheating temperature is given by\footnote{
Note that if the reheating temperature exceeds the inflaton mass, one needs to take account of the dissipation 
effect as well as non-perturbative particle production to estimate the precise reheating temperature~\cite{Mukaida:2012qn}. 
In this case thermal leptogenesis, instead of the non-thermal one, takes place. 
%For simplicity we focus on the non-thermal 
%leptogenesis by the sneutrino decay in the text, assuming that the corresponding neutrino Yukawa couplings are sufficiently small.
}
\begin{equation}
	T_{\rm R} \simeq 6\times 10^{11}\,{\rm GeV} \left( \frac{M_{N_1}}{10^{13}\,{\rm GeV}} \right)^{1/2} 
	\left( \frac{\sqrt{ \sum_{\alpha} | h_{1\alpha} |^2 }}{10^{-3}} \right).
\end{equation}
The CP violating decay of the sneutrino produces a non-zero lepton asymmetry~\cite{Murayama:1992ua, Murayama:1993em,Hamaguchi:2001gw}, and
the produced lepton asymmetry is evaluated as
\begin{equation}
	\frac{n_L}{s} \simeq 7\times 10^{-6}\left( \frac{T_{\rm R}}{10^{11}\,{\rm GeV}} \right)
	\left( \frac{m_{\nu_3}}{0.05\,{\rm eV}} \right) \delta_{\rm eff} \Delta,
\end{equation}
where $T_{\rm R}$ is the reheating temperature, $m_{\nu_3}$ is the neutrino mass indicated by the measurements 
of the atmospheric neutrinos and $\delta_{\rm eff}$ is the effective CP angle.
The dilution factor by the modulus decay is represented by $\Delta$, which will be estimated later.
The produced lepton asymmetry is converted to the baryon asymmetry through the sphaleron process.
We shall see that the entropy production by the modulus decay can be suppressed for a sufficiently heavy
modulus mass, and then the successful non-thermal leptogenesis by the decay of inflaton right-handed sneutrino is
possible.

%%%%%%%%%%%%%%%%%%%%%%%%%%%%%%%%%%%%%
\subsection{Modulus dynamics}
%%%%%%%%%%%%%%%%%%%%%%%%%%%%%%%%%%%%%

Let us study the modulus dynamics to estimate the entropy production from the modulus decay.
The minimum of the modulus $\eta$ is located at $\eta \simeq 3/4$  during inflation as  $|W| \gg |W_\phi|$.
The modulus mass at the minimum is larger than the Hubble parameter, i.e., $ |W| \gg H$, 
and therefore it is stabilized at the minimum during inflation.
After inflation, the inflaton F-term dominates over the superpotential, $|W_\phi| \gg |W|$, 
and the minimum of the modulus is shifted to $\eta \sim \sqrt{3}/4$ where the modulus mass 
is given by $|W_\phi| \sim H$.
Finally, as the inflaton oscillation amplitude decreases,  
the superpotential becomes larger than the inflaton F-term,  $|W_\phi| \ll |W| \sim m_{3/2}$, 
where  $m_{3/2}$  is the gravitino mass in the low energy.
Then the minimum again moves to $\eta \simeq 3/4$ where the modulus mass is $\simeq 6  m_{3/2}$.
In this process the modulus starts to oscillate with an amplitude of order $\eta_i \sim 0.1$.
Thus the Universe will be dominated by the modulus coherent oscillations soon after the reheating, 
and there is a cosmological moduli problem~\cite{Coughlan:1983ci,deCarlos:1993jw}.

Let us study  the modulus decay processes. To be concrete we express the modulus $z$ as
\beq
z \;\equiv\; \frac{\tau+ i a}{\sqrt{2K_{zz}}},
\eeq
where $\tau$ and $a$ are canonically normalized real and imaginary components,
and $K_{zz} = 4/3$ at the potential minimum, where
the modulus $\tau$ has a mass $m_\tau \simeq 6  m_{3/2}$
while the axion $a$ remains massless.
% It is $\tau$ that dominates the energy density of the Universe.
The modulus $\tau$ decays into a pair of gravitinos with the rate~\cite{Endo:2006zj},
\begin{align}
\Gamma(\tau  \rightarrow 2\psi_{3/2}) & \simeq \frac{1}{96 \pi} \frac{m_\tau^5}{m_{3/2}^2} 
\left(1-\frac{4 m_{3/2}^2}{m_\tau^2} \right)^\frac{3}{2}
\left(1- 6 \frac{m_{3/2}^2}{m_\tau^2} 
+ {\cal O}\left(\frac{m_{3/2}^4}{m_\tau^4}\right) \right).
\end{align}
Note that the gravitino production rate is enhanced by a factor of $m_\tau^2/m_{3/2}^2$
due to the longitudinal component. 
The modulus $\tau$ can also decay into a pair of  axions $a$ with the rate~\cite{Higaki:2013lra}
\begin{align}
\Gamma(\tau  \rightarrow 2a) & = \frac{1}{64 \pi} \frac{K_{zzz}^2}{K_{zz}^3} m_\tau^3,\\
&=\frac{1}{48 \pi} m_\tau^3,
\end{align}
where we have used $K_{zzz} = 16/9$ at the potential minimum in the second equality.
Therefore, if the modulus does not have any other interactions, it mainly decays
into gravitinos, which dominate the Universe for a while and then decay into lighter degrees of 
freedom including the standard model particles~\cite{Jeong:2012en}. 
The entropy dilution factor $\Delta (<1)$ is given by\footnote{The total e-folding number becomes
close to $50$ when there is a large entropy production by the modulus decay. }
\begin{equation}
	\Delta = {\rm min}\left[1,
	\frac{3 m_\tau T_{3/2}}{B_{3/2} m_{3/2} T_{\rm R} \eta_i^2  }
	%\frac{6M_P^2 T_z}{\eta_i^2 T_{\rm R}}
	\right] 
	\simeq {\rm min}\left[1, ~~7\times 10^{-5}\left( \frac{0.1}{\eta_i} \right)^2
	\left( \frac{m_{3/2}}{10^{9}\,{\rm GeV}} \right)^{3/2} \left(\frac{10^{11}\,{\rm GeV}}{T_{\rm R}} \right) \right],
	\label{delta}
\end{equation}
where $B_{3/2}$ is the branching fraction of the modulus decay into gravitinos, and 
$T_{3/2}$ is the decay temperature of the gravitinos. In the second equality we set $B_{3/2} \simeq 1$
and used the gravitino decay rate given by
\beq
\Gamma_{3/2} \;\simeq\; \frac{193}{384 \pi} m_{3/2}^3,
\eeq
assuming that it decays into the standard model particles and their superpartners. 
The gravitino decay temperature is estimated as
\beq
T_{3/2} \;\simeq\; 4{\rm\,TeV} \lrfp{m_{3/2}}{\GEV{9}}{\frac{3}{2}}.
\label{graT}
\eeq
Therefore, one needs a  heavy gravitino mass, $m_{3/2} \gtrsim 10^{9-10}\,{\rm GeV}$,
for successful leptogenesis. 
Note that the final baryon asymmetry becomes independent of the reheating temperature.
The abundance of axions produced by the modulus decay is diluted by the gravitino decay
and its contribution  to the effective neutrino species is given by $\Delta N_{\rm eff} \sim 0.03$.

In this minimal set-up, the SUSY breaking is not mediated to the SM sector. In particular,
there are no anomaly mediation contributions~\cite{Izawa:2010ym}.
We can generate soft SUSY breaking masses for the superpartners of the SM particles by introducing an extra
SUSY breaking sector whose effect is transmitted to the SM sector by gauge interactions. 
The soft SUSY breaking mass scale can be of order TeV, 
and some of the superparticles may be within the reach of LHC.

So far we have assumed a specific form of the K\"ahler potential, which however may be
subject to various corrections such as graviton-gravitino loops.  It is however difficult to quantify such effects
on the moduli stabilization and the contributions to the soft SUSY breaking masses
from an effective field theory point of view. In general, we expect that the soft SUSY breaking masses
for sfermions will be a few orders of magnitude smaller than the gravitino mass, if such corrections are
induced radiatively.  Such heavy sfermion mass, especially the stop mass, of order $\GEV{6-7}$ is consistent with the SM-like Higgs
boson of mass near 126\,GeV~\cite{Aad:2012tfa,Chatrchyan:2012ufa}.
On the other hand, unless $z$ has a direct coupling to the SM gauge fields (which will be
considered below), the gaugino mass remains significantly suppressed and it 
arises only at the two-loop level and given by $\sim m_{3/2}^3$~\cite{ArkaniHamed:2004yi}, where
we have neglected loop factors. Therefore we need to invoke an additional SUSY breaking
and its mediation to the visible sector, in order to generate a sizable gluino mass $\gtrsim$\,TeV. The resultant 
soft mass spectrum resembles that in split SUSY~\cite{ArkaniHamed:2004yi} or 
pure gravity mediation~\cite{Ibe:2011aa,Ibe:2012hu} scenarios.

Another possible extension is to introduce  couplings of $z$.   Let us here briefly discuss what happens if $z$
is coupled to the SM gauge sector.
%Note that the axion $a$ is massless in this case.
%If there is entropy production by some other fields, successful cosmology may be possible.\footnote{
%For an appropriate amount of entropy production, it is possible to realize $\Delta N_{\rm eff} \sim 0.3$ 
%suggested by the recent observations~\cite{Ade:2013rta}.
%}
%To avoid the overproduction of axions, 
We introduce the following coupling to the gauge bosons:
\begin{equation}
	\mathcal L = \int d^2\theta \frac{z}{M} W^\alpha W_\alpha+ {\rm h.c.}
	\label{zWW}
\end{equation}
where $M$ is an effective cutoff scale and $W^\alpha$ denotes the
SM gauge superfield. We assume that the gauge superfields are canonically normalized,
which is not modified as $\la z \ra \ll M$. 
%Note that the interaction (\ref{zWW}) gives the gaugino masses since $z$ dominantly breaks SUSY.
The gaugino mass is generated by the above interaction,
\begin{equation}
	m_\lambda =2 K^{z\bar z}K_{\bar z} \frac{m_{3/2}}{M} = \frac{3m_{3/2}}{M}.
	% \frac{1}{2{\rm Re}(z)} K^{z\bar z}K_{\bar z} \frac{m_{3/2}}{M} = \frac{4m_{3/2}}{M}. 
\end{equation}
For instance, the gaugino mass of ${\cal O}(1)$\,TeV is generated for $M \simeq 10^6 M_p$
and $m_{3/2} \simeq \GEV{9}$. In the minimal set-up,
%The sfermions do not obtain the masses of order of the gravitino in our setup, and their 
the sfermion masses dominantly come from the renormalization group evolution effect as in the gaugino
 mediation model~\cite{Inoue:1991rk,Kaplan:1999ac}.
Thus the squark/slepton masses are suppressed by a loop factor compared with the gauginos.
For a sufficiently large $M$, the soft SUSY breaking masses can be of ${\cal O}(1-10)$\,TeV.
The 126\,GeV Higgs boson mass can be explained in such a setup~\cite{Moroi:2012kg}.

The above coupling induces the decay of modulus into the gauge boson as
\begin{equation}
	\Gamma(z\to A_\mu A_\mu) \simeq \frac{3N_g}{32\pi} \frac{m_\tau^3}{M^2},  \label{ztogg}
\end{equation}
where $N_g$ is the number of gauge bosons, and we have $N_g = 12$ in the SM.
The modulus decays also into gauginos with a similar rate. 
The partial decay rate into the SM gauge sector is smaller than that into gravitinos unless
$M$ is much smaller than the Planck scale, and the above estimate on the entropy dilution 
factor remains almost unchanged. 
%
%Then the branching fraction into axions is $4\%$, and we can avoid the axion overproduction.
%The dilution factor for the lepton asymmetry is then evaluated as
%%%
%\begin{equation}
%	\Delta = {\rm min}\left[1,\frac{6M_P^2 T_z}{\eta_i^2 T_{\rm R}}\right] 
%	\simeq {\rm min}\left[1, ~~4\times 10^{-2}\left( \frac{0.1}{\eta_i} \right)^2
%	\left( \frac{m_\tau}{10^{10}\,{\rm GeV}} \right)^{3/2} \left(\frac{10^{11}\,{\rm GeV}}{T_{\rm R}} \right) \right],
%	\label{delta}
%\end{equation}
%%%
%where $T_z$ is the modulus decay temperature.
%Therefore, the successful leptogenesis requires $m_\tau\sim m_{3/2} \gtrsim 10^{9-10}\,{\rm GeV}$.

The axion $a$ becomes the QCD axion as it acquires a mass from the QCD instanton effect through 
\EQ{zWW}. The axion decay constant $f_a$ is related to the effective cut-off $M$ as
\beq
f_a \;=\; \frac{\sqrt{2 K_{z {\bar z}}}}{32 \pi^2} M,
\eeq
and the axion mass is given by
\beq
m_a \;\simeq\; 5 \times 10^{-16}{\rm\,eV} \lrf{10^6 M_p}{M}.
\eeq
However the axion isocurvature perturbation becomes too large in this case 
no matter how the initial misalignment angle is tuned, because of the high inflation scale~\cite{Kawasaki:2008sn,Hikage:2012be}.
One solution to this problem is to introduce a coupling of $z$ to another hidden strong gauge group so that
the axion gets a heavy mass of $\sim \Lambda^2/M$ during inflation, where $\Lambda$ is the dynamical scale of the hidden 
gauge group.  If the hidden-gauge group remains strongly coupled in the low-energy, 
 the axion does not solve the strong CP problem, and even if it is produced by the coherent oscillations, it
decays into  the SM gauge bosons, thus avoiding the isocurvature constraint. Note that
one can choose the value of $\Lambda$ so that it does not modify the moduli stabilization significantly. 
Alternatively, if the hidden-gauge group becomes weakly-coupled in the low-energy somehow by e.g. non-trivial
dynamics of a dilation field or hidden Higgs fields~\cite{Dvali:1995ce}, the axion may be able to solve the strong CP problem, avoiding
the isocurvature constraint~\cite{Jeong:2013xta}. 

%For $\Lambda \sim {\cal O}(10^{13})$\,GeV, the axion mass $m_a$ is about  ${\cal O}(1-10\%)$ of the 
%modulus mass. As a result,  the entropy dilution factor is still given by \EQ{delta}, and
%the stabilization of $\tau$ is also not modified significantly. 

%On the other hand, the imaginary component of $z$, which we denote by $a$, is 

%Thus it acts as the QCD axion to solve the strong CP problem.
%The modulus decays into its axionic component with the decay rate~\cite{Higaki:2013lra}
%%
%\begin{equation}
%	\Gamma(z\to aa) \simeq \frac{289}{3072\pi}m_z^3.
%\end{equation}
%%
%Thus the branching ratio into the axion pair, $B_a$, is around $B_a \sim 0.2$.
%It yields the axionic dark radiation $\Delta N_{\rm eff} \sim 1$ in terms of the extra effective number of relativistic species,
%which is observable by future experiments.

Lastly let us comment on the lightest SUSY particle (LSP). The SUSY particles are produced by the gravitino decays.
Since the gravitino decay temperature is higher than TeV for $m_{3/2} \gtrsim \GEV{9}$ (see \EQ{graT}),
the LSPs are thermalized if their mass is of order 100 to 1000\,GeV. Then a right amount of dark matter can be explained by
the thermal relic of the LSPs. On the other hand, if the LSP mass is much larger than TeV, the thermal relic abundance
likely exceeds the observed dark matter abundance, and one would need to introduce a small amount of R-parity violation.

%%%%%%%%%%%%%%%%%%%%%%%%%%%%%%%%%%%%%
\section{Discussion and Conclusions}
%%%%%%%%%%%%%%%%%%%%%%%%%%%%%%%%%%%%%

In this paper we have revisited the chaotic inflation model in which the lightest right-handed sneutrino
plays the role of the inflaton.  The model predicts a rather large tensor-to-scalar ratio, which is within the 
reach of the  future  and on-going B-mode  search experiments. Furthermore, the inflaton naturally
reheats the SM particles and non-thermal leptogenesis takes place naturally. 

We have also embedded the right-handed
sneutrino inflation model in a supergravity framework, and shown that 
the inflaton dynamics is same as in the global SUSY case for a certain class of the K\"ahler potential.  
The price we have to pay for obtaining the inflaton potential as in the global SUSY is the existence
of a modulus field, which causes a cosmological moduli problem. 
We have  shown that
 that the gravitino mass should be sufficiently heavy, i.e. $m_{3/2} \gtrsim \GEV{9}$, 
for successful leptogenesis, since otherwise the modulus (and gravitino) decay would dilute the baryon asymmetry
too much. 

The soft mass spectrum in the visible sector depends on the precise form of the K\"ahler potential.
As long as it is given by \EQ{Km} (a more general form will be discussed in Appendix), 
the structure of the visible sector is essentially same as in the global SUSY, and the SUSY
breaking effect is not mediated to the visible sector. In particular, there is no anomaly mediation contribution.
In this case we need to invoke an additional SUSY breaking and its mediation mechanism to the visible sector.
We however note that the K\"ahler potential could receive various corrections such as graviton-gravitino and 
moduli loops. In this case, we expect that sfermions obtain a SUSY breaking mass  a few orders of magnitude smaller 
than the gravitino mass, while the gaugino mass remains significantly suppressed, which requires an additional
SUSY breaking and its mediation mechanism. The resultant soft SUSY mass spectrum will be similar to those in the
split SUSY and pure gravity mediation scenarios.

%Second, SUSY particles are as heavy as $\sim 10^{7-8}\,$GeV, which are beyond the reach of the LHC experiment.

Some comments are in order.
We have dropped higher order terms in (\ref{W}). This is justified as the inflaton
has a mass about one order of magnitude smaller than the naively expected value.
Otherwise, higher order terms are generically non-negligible where the first term
and the second term in (\ref{W}) become comparable to each other. That said,
it is in principle possible that the higher order terms modify the inflaton potential.
It may be possible to lift the inflaton potential  at large field values so that there
is no local minimum. In this case, there will be no problem of choosing the initial
position of the inflaton near the local maximum. In particular,  the inflaton potential
can be flatter at large field values, which will lead to a smaller  tensor-to-scalar ratio 
in better agreement with the observation.

%%%%%%%%%%%%%%%%%%%%%%%%%%%%%%%%%%%%%%%%%%%%
\section*{Acknowledgments}
%%%%%%%%%%%%%%%%%%%%%%%%%%%%%%%%%%%%%%%%%%%%
We are grateful to John Ellis for fruitful discussion in the early stage 
of the project, and we  thank Hitoshi Murayama for discussion. 
This work was supported by the Grant-in-Aid for Scientific Research on
Innovative Areas (No. 21111006  [KN and FT],  No.23104008 [FT], No.24111702 [FT]),
Scientific Research (A) (No. 22244030 [KN and FT], 21244033 [FT], 22244021 [TTY]),  JSPS Grant-in-Aid for
Young Scientists (B) (No.24740135) [FT], and Inoue Foundation for Science [FT].  This work was also
supported by World Premier International Center Initiative (WPI Program), MEXT, Japan.

\appendix

%%%%%%%%%%%%%%%%%%%%%%%%%%%%%%%%%%%%%%%%%%%%
\section{Condition for successful chaotic inflation} \label{sec:app}
%%%%%%%%%%%%%%%%%%%%%%%%%%%%%%%%%%%%%%%%%%%%

In this Appendix we derive conditions for successful chaotic inflation with $|\phi| \gg M_P$.
Let us consider the following K\"ahler potential and superpotential:
\begin{gather}
	K = f(\eta), \\
	W = W (\phi),
\end{gather}
where
\begin{equation}
	\eta = z + z^\dagger + c |\phi|^2,
\end{equation}
with a numerical constant $c$.
It exhibits the Heisenberg symmetry for $c=1$.
The kinetic term is given by
\begin{equation}
	\mathcal L_{\rm kin} = \frac{f''}{4}\left[ (\partial \eta)^2 + I_\mu I^\mu \right] + cf' |\partial \phi|^2,
\end{equation}
where $I_\mu = i\partial_\mu(z-z^\dagger) + ic(\phi^\dagger \partial_\mu\phi -\phi \partial_\mu\phi^\dagger)$.
The scalar potential is given by
\begin{equation}
	V = e^f \left[  \frac{1}{cf'}|W_\phi|^2 + \left(\frac{f'^2}{f''}-3\right)|W|^2  \right],
	\label{Vsugra2}
\end{equation}
where the prime denotes the derivative with respect to $\eta$.
For chaotic inflation to happen, we demand the following relations at $\eta = \eta_{\rm min}$:
\begin{gather}
	F(\eta_{\rm min}) \equiv \left[ \frac{f'^2}{f''} - 3 \right]_{\eta = \eta_{\rm min}} = 0, \nonumber \\
	F'(\eta_{\rm min}) = \left[\frac{f'(2f^{''2}-f'f''')}{f^{''2}} \right]_{\eta = \eta_{\rm min}}= 0,  \label{F'} \\
	F''(\eta_{\rm min}) = \left[ \frac{f'}{f^{''2}}(3f'' f''' - f' f'''') \right]_{\eta = \eta_{\rm min}}> 0 \nonumber.
\end{gather}
If these are satisfied, $\eta$ is stabilized at $\eta = \eta_{\rm min}$, where the dangerous second term in (\ref{Vsugra2}) vanishes.
Then the potential for $\phi$ may resemble that in the global SUSY case even for $|\phi| \gg 1$.

\begin{figure}
\begin{center}
\includegraphics[scale=1.2]{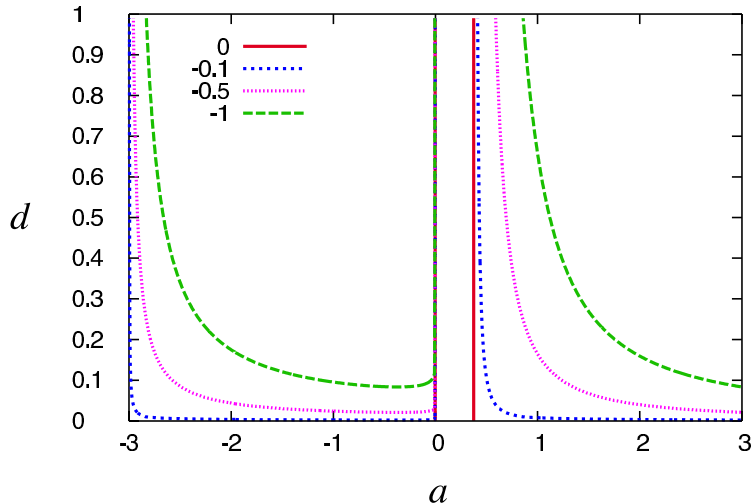}
\caption{ 
	Contours of $b$ satisfying conditions (\ref{cond1})-(\ref{cond2}) on $(a,d)$-plane.
}
\label{fig:ad}
\end{center}
\end{figure}

To be more concrete, let us assume the following form:
\begin{equation}
	f(\eta) = a\ln \eta + b\eta + d\eta^2,
\end{equation}
with numerical coefficients $a,b$ and $d$.
From Eqs.~(\ref{F'}), we find 
\begin{equation}
	2d\eta_{\rm min}^2 = (3a^2)^{1/3} + a,
	\label{cond1}
\end{equation}
and
\begin{equation}
	4d^2\eta_{\rm min}^3 -6ad \eta_{\rm min} - ab = 0.
	\label{cond2}
\end{equation}
We also have
\begin{equation}
	F''(\eta_{\rm min}) = 6\left(4d + \frac{b}{4\eta_{\rm min}} \right).
\end{equation}
For example, if we take $b=0$ and $d=1$, we find $a=3/8$ and $\eta_{\rm min} = 3/4$
as found in Ref.~\cite{Murayama:1993xu}.
If we take $b=0$ and $d=0$, we find $a=-3$ as in the no-scale form, 
although $\eta$ is massless and not stabilized since $F''(\eta_{\rm min})=0$ in this limit.
Fig.~\ref{fig:ad} shows contours of $b$ satisfying conditions (\ref{cond1})-(\ref{cond2}) on $(a,d)$-plane.
It is checked that $F''(\eta_{\rm min}) > 0$ for all the parameter ranges.

%%%%%%%%%%%%%%%%
%%%%%%%%%%%%%%%%

%%%%%%%%%%%%%%%%%%%%%%%%%%%%%%%%%%%%%

%%%%%%%%%%%%%%%%%%%%%%%%%%%%%%%%%%%%%


\begin{thebibliography}{99}
%%%%%%%%%%%%%%%%%%%%%%%%%%%%%%%%%%%%%
    

  
   %\cite{Ade:2013rta}
\bibitem{Ade:2013rta} 
  P.~A.~R.~Ade {\it et al.}  [ Planck Collaboration],
  %``Planck 2013 results. XXII. Constraints on inflation,''
  arXiv:1303.5082 [astro-ph.CO].
  %%CITATION = ARXIV:1303.5082;%%

  %\cite{Guth:1980zm}
\bibitem{Guth:1980zm}
  A.~H.~Guth,
  %``The Inflationary Universe: A Possible Solution to the Horizon and Flatness Problems,''
  Phys.\ Rev.\  D {\bf 23}, 347-356 (1981);
  %\cite{Starobinsky:1980te}
%\bibitem{Starobinsky:1980te}
A.~A.~Starobinsky,
%``A New Type of Isotropic Cosmological Models without Singularity,''
Phys.\ Lett.\ B {\bf 91} (1980) 99;
%%CITATION = PHLTA,B91,99;%%
%\cite{Sato:1980yn}
%\bibitem{Sato:1980yn}
  K.~Sato,
  %``First Order Phase Transition of a Vacuum and Expansion of the Universe,''
  Mon.\ Not.\ Roy.\ Astron.\ Soc.\  {\bf 195}, 467-479 (1981).
  
  %\cite{Linde:1981mu}
\bibitem{Linde:1981mu}
A.~D.~Linde,
%``A New Inflationary Universe Scenario: a Possible Solution of the Horizon, Flatness, Homogeneity, Isotropy and Primordial Monopole Problems,''
Phys.\ Lett.\ B {\bf 108} (1982) 389;
%%CITATION = PHLTA,B108,389;%%
%\cite{Albrecht:1982wi}
%\bibitem{Albrecht:1982wi} 
  A.~Albrecht and P.~J.~Steinhardt,
  %``Cosmology for Grand Unified Theories with Radiatively Induced Symmetry Breaking,''
  Phys.\ Rev.\ Lett.\  {\bf 48}, 1220 (1982).
  %%CITATION = PRLTA,48,1220;%%
  
     %\cite{Linde:1983gd}
\bibitem{Linde:1983gd}
  A.~D.~Linde,
  %``Chaotic Inflation,''
  Phys.\ Lett.\ B {\bf 129}, 177 (1983).
  %%CITATION = PHLTA,B129,177;%%   
  
  %\cite{Croon:2013ana}
\bibitem{Croon:2013ana} 
  D.~Croon, J.~Ellis and N.~E.~Mavromatos,
  %``Wess-Zumino Inflation in Light of Planck,''
  Physics Letters B {\bf 724}, , 165 (2013)
  [arXiv:1303.6253 [astro-ph.CO]].
  %%CITATION = ARXIV:1303.6253;%%
  
  %\cite{Nakayama:2013jka}
\bibitem{Nakayama:2013jka} 
  K.~Nakayama, F.~Takahashi and T.~T.~Yanagida,
  %``Polynomial Chaotic Inflation in the Planck Era,''
  Phys.\  Lett.\  B725, {\bf 111} (2013)
  [arXiv:1303.7315 [hep-ph]];
  %%CITATION = ARXIV:1303.7315;%%
  %\cite{Nakayama:2013txa}
%\bibitem{Nakayama:2013txa} 
  %K.~Nakayama, F.~Takahashi and T.~T.~Yanagida,
  %``Polynomial Chaotic Inflation in Supergravity,''
  JCAP {\bf 1308}, 038 (2013)
  [arXiv:1305.5099 [hep-ph]].
  %%CITATION = ARXIV:1305.5099,;%%
  
%\cite{Ellis:2013xoa}
\bibitem{Ellis:2013xoa} 
  J.~Ellis, D.~V.~Nanopoulos and K.~A.~Olive,
  %``No-Scale Supergravity Realization of the Starobinsky Model of Inflation,''
  Phys.\ Rev.\ Lett.\  {\bf 111}, 111301 (2013)
  [arXiv:1305.1247 [hep-th]].
  %%CITATION = ARXIV:1305.1247;%%
  %\cite{Ellis:2013nxa}
%\cite{Ellis:2013nxa}
%\bibitem{Ellis:2013nxa} 
%  J.~Ellis, D.~V.~Nanopoulos and K.~A.~Olive,
  %``Starobinsky-like Inflationary Models as Avatars of No-Scale Supergravity,''
  JCAP {\bf 1310}, 009 (2013)
  [arXiv:1307.3537 [hep-th]];
  %%CITATION = ARXIV:1307.3537;%%
  %20 citations counted in INSPIRE as of 09 Dec 2013
%\cite{Ellis:2013nka}
%\bibitem{Ellis:2013nka} 
%  J.~Ellis, D.~V.~Nanopoulos and K.~A.~Olive,
  %``A No-Scale Framework for Sub-Planckian Physics,''
  arXiv:1310.4770 [hep-ph].
  %%CITATION = ARXIV:1310.4770;%%
  %4 citations counted in INSPIRE as of 09 Dec 2013


  %\cite{Kallosh:2013pby}
\bibitem{Kallosh:2013pby} 
  R.~Kallosh and A.~Linde,
  %``Superconformal generalization of the chaotic inflation model $\frac{\lambda}{4} \phi^{4} - \frac{\xi}{2} \phi^{2}R$,''
  JCAP {\bf 1306}, 027 (2013)
  [arXiv:1306.3211 [hep-th]];
  %%CITATION = ARXIV:1306.3211;%%
  %\cite{Kallosh:2013lkr}
%\bibitem{Kallosh:2013lkr} 
  %R.~Kallosh and A.~Linde,
  %``Superconformal generalizations of the Starobinsky model,''
  JCAP {\bf 1306}, 028 (2013)
  [arXiv:1306.3214 [hep-th]].
  %%CITATION = ARXIV:1306.3214;%%
  
  %\cite{Destri:2007pv}
\bibitem{Destri:2007pv} 
  C.~Destri, H.~J.~de Vega and N.~G.~Sanchez,
  %``MCMC analysis of WMAP3 and SDSS data points to broken symmetry inflaton potentials and provides a lower bound on the tensor to scalar ratio,''
  Phys.\ Rev.\ D {\bf 77}, 043509 (2008)
  [astro-ph/0703417].
  %%CITATION = ASTRO-PH/0703417;%%
  %34 citations counted in INSPIRE as of 01 Apr 2013
  
  %\cite{Linde:1983fq}
  \bibitem{Linde:1983fq}   
  A.~D.~Linde, 
   %``Supergravity And Inflationary Universe. (in Russian),''
  Pisma Zh.\ Eksp.\ Teor.\ Fiz.\  {\bf 37}, 606 (1983)
  [JETP Lett.\  {\bf 37}, 724 (1983)];
  %%CITATION = ZFPRA,37,606;%%
  %11 citations counted in INSPIRE as of 31 Mar 2013
%\cite{Linde:1984cd}
%\bibitem{Linde:1984cd}   
%A.~D.~Linde,  %``Primordial Inflation Without Primordial Monopoles,''
  Phys.\ Lett.\ B {\bf 132}, 317 (1983).
  %%CITATION = PHLTA,B132,317;%%
  %65 citations counted in INSPIRE as of 31 Mar 2013
    
    %\cite{Kallosh:2007wm}
\bibitem{Kallosh:2007wm} 
  R.~Kallosh and A.~D.~Linde,
  %``Testing String Theory with CMB,''
  JCAP {\bf 0704}, 017 (2007)
  [arXiv:0704.0647 [hep-th]].
  %%CITATION = ARXIV:0704.0647;%%
  %56 citations counted in INSPIRE as of 02 Apr 2013      

%\cite{Murayama:1992ua}
\bibitem{Murayama:1992ua} 
  H.~Murayama, H.~Suzuki, T.~Yanagida and J.~'i.~Yokoyama,
  %``Chaotic inflation and baryogenesis by right-handed sneutrinos,''
  Phys.\ Rev.\ Lett.\  {\bf 70}, 1912 (1993).
  %%CITATION = PRLTA,70,1912;%%
  %187 citations counted in INSPIRE as of 09 Jan 2014

    %\cite{Murayama:1993xu}
\bibitem{Murayama:1993xu} 
  H.~Murayama, H.~Suzuki, T.~Yanagida and J.~'i.~Yokoyama,
  %``Chaotic inflation and baryogenesis in supergravity,''
  Phys.\ Rev.\ D {\bf 50}, 2356 (1994)
  [hep-ph/9311326].
  %%CITATION = HEP-PH/9311326;%%

%\cite{Ellis:2003sq}
\bibitem{Ellis:2003sq} 
  J.~R.~Ellis, M.~Raidal and T.~Yanagida,
  %``Sneutrino inflation in the light of WMAP: Reheating, leptogenesis and flavor violating lepton decays,''
  Phys.\ Lett.\ B {\bf 581}, 9 (2004)
  [hep-ph/0303242].
  %%CITATION = HEP-PH/0303242;%%
  %91 citations counted in INSPIRE as of 14 Nov 2013
  
  %\cite{Antusch:2009ty}
\bibitem{Antusch:2009ty} 
  S.~Antusch, M.~Bastero-Gil, K.~Dutta, S.~F.~King and P.~M.~Kostka,
  %``Chaotic Inflation in Supergravity with Heisenberg Symmetry,''
  Phys.\ Lett.\ B {\bf 679}, 428 (2009)
  [arXiv:0905.0905 [hep-th]].
  %%CITATION = ARXIV:0905.0905;%%
  %15 citations counted in INSPIRE as of 26 Jan 2014

   \bibitem{seesaw}
T.~Yanagida, in Proceedings of the {\it{``Workshop on the Unified Theory and
 the Baryon Number in the Universe''}}, Tsukuba, Japan, Feb. 13-14, 1979, edited by
O.~Sawada and A.~Sugamoto, KEK report KEK-79-18, p. 95,
and {\it{``Horizontal Symmetry And Masses Of Neutrinos''
}}, Prog. Theor. Phys. {\bf{64}} (1980) 1103;
M.~Gell-Mann, P.~Ramond and R.~Slansky, in {\it{``Supergravity''}}
 (North-Holland, Amsterdam, 1979) {\it{eds}}. D.~Z.~Freedom and P.~van
Nieuwenhuizen, Print-80-0576 (CERN);
%\cite{Glashow:1979nm}
%\bibitem{Glashow:1979nm} 
  S.~L.~Glashow,
  %``The Future of Elementary Particle Physics,''
  NATO Adv.\ Study Inst.\ Ser.\ B Phys.\  {\bf 59}, 687 (1980);
  %%CITATION = NASBD,59,687;%%
  %173 citations counted in INSPIRE as of 08 Jan 2014
  %\cite{Mohapatra:1979ia}
%\bibitem{Mohapatra:1979ia} 
  R.~N.~Mohapatra and G.~Senjanovic,
  %``Neutrino Mass and Spontaneous Parity Violation,''
  Phys.\ Rev.\ Lett.\  {\bf 44}, 912 (1980);
  %%CITATION = PRLTA,44,912;%%
  %3279 citations counted in INSPIRE as of 08 Jan 2014
see also   P.~Minkowski,  Phys.\ Lett.\  B {\bf 67}, 421 (1977).

%\cite{Hall:1999sn}
\bibitem{Hall:1999sn}
  L.~J.~Hall, H.~Murayama and N.~Weiner,
  %``Neutrino mass anarchy,''
  Phys.\ Rev.\ Lett.\  {\bf 84}, 2572 (2000)
  [hep-ph/9911341].
  %%CITATION = HEP-PH/9911341;%%
  %194 citations counted in INSPIRE as of 30 Aug 2013

  %\cite{Haba:2000be}
\bibitem{Haba:2000be}
  N.~Haba and H.~Murayama,
  %``Anarchy and hierarchy,''
  Phys.\ Rev.\ D {\bf 63}, 053010 (2001)
  [hep-ph/0009174].
  %%CITATION = HEP-PH/0009174;%%
  %136 citations counted in INSPIRE as of 30 Aug 2013


\bibitem{Fukugita:1986hr}
  M.~Fukugita and T.~Yanagida,
  %``Baryogenesis Without Grand Unification,''
  Phys.\ Lett.\ B {\bf 174}, 45 (1986).
  
  %\cite{Jeong:2012zj}
\bibitem{Jeong:2012zj} 
  K.~S.~Jeong and F.~Takahashi,
  %``Anarchy and Leptogenesis,''
  JHEP {\bf 1207}, 170 (2012)
  [arXiv:1204.5453 [hep-ph]].
  %%CITATION = ARXIV:1204.5453;%%
  %2 citations counted in INSPIRE as of 18 Oct 2013

  

  %\cite{Mukaida:2012qn}
\bibitem{Mukaida:2012qn} 
  K.~Mukaida and K.~Nakayama,
  %``Dynamics of oscillating scalar field in thermal environment,''
  JCAP {\bf 1301}, 017 (2013)
  [arXiv:1208.3399 [hep-ph]];
  %%CITATION = ARXIV:1208.3399;%%
  %\cite{Mukaida:2012bz}
%\bibitem{Mukaida:2012bz} 
  %K.~Mukaida and K.~Nakayama,
  %``Dissipative Effects on Reheating after Inflation,''
  JCAP {\bf 1303}, 002 (2013)
  [arXiv:1212.4985 [hep-ph]].
  %%CITATION = ARXIV:1212.4985;%%
  
%  %\cite{Lazarides:1991wu}
%\bibitem{Lazarides:1991wu} 
%  G.~Lazarides and Q.~Shafi,
%  %``Origin of matter in the inflationary cosmology,''
%  Phys.\ Lett.\ B {\bf 258}, 305 (1991).
%  %%CITATION = PHLTA,B258,305;%%
%
%%\cite{Asaka:1999yd}
%\bibitem{Asaka:1999yd}
%  T.~Asaka, K.~Hamaguchi, M.~Kawasaki, T.~Yanagida,
%  %``Leptogenesis in inflaton decay,''
%  Phys.\ Lett.\  {\bf B464}, 12-18 (1999)
%  [hep-ph/9906366];
%  %\cite{Asaka:1999jb}
%%\bibitem{Asaka:1999jb}
%  %T.~Asaka, K.~Hamaguchi, M.~Kawasaki, T.~Yanagida,
%  %``Leptogenesis in inflationary Universe,''
%  Phys.\ Rev.\  {\bf D61}, 083512 (2000)
%  [hep-ph/9907559].
  
%\cite{Murayama:1993em}
\bibitem{Murayama:1993em} 
  H.~Murayama and T.~Yanagida,
  %``Leptogenesis in supersymmetric standard model with right-handed neutrino,''
  Phys.\ Lett.\ B {\bf 322}, 349 (1994)
  [hep-ph/9310297].
  %%CITATION = HEP-PH/9310297;%%
  %147 citations counted in INSPIRE as of 09 Jan 2014  
  
    %\cite{Hamaguchi:2001gw}
\bibitem{Hamaguchi:2001gw} 
  K.~Hamaguchi, H.~Murayama and T.~Yanagida,
  %``Leptogenesis from N dominated early universe,''
  Phys.\ Rev.\ D {\bf 65}, 043512 (2002)
  [hep-ph/0109030].
  %%CITATION = HEP-PH/0109030;%%
  
  
  %\cite{Coughlan:1983ci}
\bibitem{Coughlan:1983ci}
  G.~D.~Coughlan, W.~Fischler, E.~W.~Kolb, S.~Raby, G.~G.~Ross,
  %``Cosmological Problems for the Polonyi Potential,''
  Phys.\ Lett.\ B {\bf 131}, 59 (1983);
%\cite{Ellis:1986zt}
%\bibitem{Ellis:1986zt}
  J.~R.~Ellis, D.~V.~Nanopoulos, M.~Quiros,
  %``On the Axion, Dilaton, Polonyi, Gravitino and Shadow Matter Problems in Supergravity and Superstring Models,''
  Phys.\ Lett.\ B {\bf 174}, 176 (1986);
  %\cite{Goncharov:1984qm}
%\bibitem{Goncharov:1984qm}
  A.~S.~Goncharov, A.~D.~Linde, M.~I.~Vysotsky,
  %``Cosmological Problems For Spontaneously Broken Supergravity,''
  Phys.\ Lett.\ B {\bf 147}, 279 (1984).
%

%\cite{deCarlos:1993jw}
\bibitem{deCarlos:1993jw}
  B.~de Carlos, J.~A.~Casas, F.~Quevedo and E.~Roulet,
  %``Model independent properties and cosmological implications of the dilaton and moduli sectors of 4-d strings,''  
Phys.\ Lett.\ B {\bf 318}, 447 (1993)  [hep-ph/9308325];
%%CITATION = HEP-PH/9308325;%%
%\cite{Banks:1993en}
%\bibitem{Banks:1993en} 
  T.~Banks, D.~B.~Kaplan and A.~E.~Nelson,
  %``Cosmological implications of dynamical supersymmetry breaking,''　
  Phys.\ Rev.\ D {\bf 49}, 779 (1994)
  [hep-ph/9308292].
  %%CITATION = HEP-PH/9308292;%%

     %\cite{Endo:2006zj}
\bibitem{Endo:2006zj}
  M.~Endo, K.~Hamaguchi, F.~Takahashi,
  %``Moduli-induced gravitino problem,''
  Phys.\ Rev.\ Lett.\  {\bf 96}, 211301 (2006).
  [hep-ph/0602061];
  %\cite{Endo:2006tf}
%\bibitem{Endo:2006tf}
  %M.~Endo, K.~Hamaguchi, F.~Takahashi,
  %``Moduli/Inflaton Mixing with Supersymmetry Breaking Field,''
  Phys.\ Rev.\  {\bf D74}, 023531 (2006).
  [hep-ph/0605091];
  %\cite{Nakamura:2006uc}
%\bibitem{Nakamura:2006uc}
  S.~Nakamura, M.~Yamaguchi,
  %``Gravitino production from heavy moduli decay and cosmological moduli problem revived,''
  Phys.\ Lett.\  {\bf B638}, 389-395 (2006).
  [hep-ph/0602081];
  %\cite{Dine:2006ii}
%\bibitem{Dine:2006ii}
  M.~Dine, R.~Kitano, A.~Morisse, Y.~Shirman,
  %``Moduli decays and gravitinos,''
  Phys.\ Rev.\  {\bf D73}, 123518 (2006).
  [hep-ph/0604140].

%\cite{Higaki:2013lra}
\bibitem{Higaki:2013lra} 
  T.~Higaki, K.~Nakayama and F.~Takahashi,
  %``Moduli-Induced Axion Problem,''
  JHEP {\bf 1307}, 005 (2013)
  [arXiv:1304.7987 [hep-ph]].
  %%CITATION = ARXIV:1304.7987;%%
  %9 citations counted in INSPIRE as of 18 Oct 2013
  
  %\cite{Jeong:2012en}
\bibitem{Jeong:2012en} 
  K.~S.~Jeong and F.~Takahashi,
  %``A Gravitino-rich Universe,''
  JHEP {\bf 1301}, 173 (2013)
  [arXiv:1210.4077 [hep-ph]].
  %%CITATION = ARXIV:1210.4077;%%
  %4 citations counted in INSPIRE as of 29 Oct 2013
  
  %\cite{Izawa:2010ym}
\bibitem{Izawa:2010ym} 
  K.~-I.~Izawa, T.~Kugo and T.~T.~Yanagida,
  %``Gravitational Supersymmetry Breaking,''
  Prog.\ Theor.\ Phys.\  {\bf 125}, 261 (2011)
  [arXiv:1008.4641 [hep-ph]].
  %%CITATION = ARXIV:1008.4641;%%
  %4 citations counted in INSPIRE as of 07 Nov 2013
  
  %\cite{Aad:2012tfa}
\bibitem{Aad:2012tfa} 
  G.~Aad {\it et al.}  [ATLAS Collaboration],
  %``Observation of a new particle in the search for the Standard Model Higgs boson with the ATLAS detector at the LHC,''
  Phys.\ Lett.\ B {\bf 716}, 1 (2012)
  [arXiv:1207.7214 [hep-ex]].
  %%CITATION = ARXIV:1207.7214;%%
  %1860 citations counted in INSPIRE as of 06 Nov 2013
  
  %\cite{Chatrchyan:2012ufa}
\bibitem{Chatrchyan:2012ufa} 
  S.~Chatrchyan {\it et al.}  [CMS Collaboration],
  %``Observation of a new boson at a mass of 125 GeV with the CMS experiment at the LHC,''
  Phys.\ Lett.\ B {\bf 716}, 30 (2012)
  [arXiv:1207.7235 [hep-ex]].
  %%CITATION = ARXIV:1207.7235;%%
  %1840 citations counted in INSPIRE as of 06 Nov 2013
  
  %\cite{ArkaniHamed:2004yi}
\bibitem{ArkaniHamed:2004yi} 
  N.~Arkani-Hamed, S.~Dimopoulos, G.~F.~Giudice and A.~Romanino,
  %``Aspects of split supersymmetry,''
  Nucl.\ Phys.\ B {\bf 709}, 3 (2005)
  [hep-ph/0409232].
  %%CITATION = HEP-PH/0409232;%%
  %379 citations counted in INSPIRE as of 06 Nov 2013
  
  
%\cite{Ibe:2011aa}
\bibitem{Ibe:2011aa} 
  M.~Ibe and T.~T.~Yanagida,
  %``The Lightest Higgs Boson Mass in Pure Gravity Mediation Model,''
  Phys.\ Lett.\ B {\bf 709}, 374 (2012)
  [arXiv:1112.2462 [hep-ph]].
  %%CITATION = ARXIV:1112.2462;%%
  %64 citations counted in INSPIRE as of 06 Nov 2013
  
  %\cite{Ibe:2012hu}
\bibitem{Ibe:2012hu} 
  M.~Ibe, S.~Matsumoto and T.~T.~Yanagida,
  %``Pure Gravity Mediation with m_{3/2} = 10-100TeV,''
  Phys.\ Rev.\ D {\bf 85}, 095011 (2012)
  [arXiv:1202.2253 [hep-ph]].
  %%CITATION = ARXIV:1202.2253;%%
  %51 citations counted in INSPIRE as of 06 Nov 2013
  
  %\cite{Inoue:1991rk}
\bibitem{Inoue:1991rk} 
  K.~Inoue, M.~Kawasaki, M.~Yamaguchi and T.~Yanagida,
  %``Vanishing squark and slepton masses in a class of supergravity models,''
  Phys.\ Rev.\ D {\bf 45}, 328 (1992).
  %%CITATION = PHRVA,D45,328;%%
  
  %\cite{Kaplan:1999ac}
\bibitem{Kaplan:1999ac} 
  D.~E.~Kaplan, G.~D.~Kribs and M.~Schmaltz,
  %``Supersymmetry breaking through transparent extra dimensions,''
  Phys.\ Rev.\ D {\bf 62}, 035010 (2000)
  [hep-ph/9911293];
  %%CITATION = HEP-PH/9911293;%%
  %\cite{Chacko:1999mi}
%\bibitem{Chacko:1999mi} 
  Z.~Chacko, M.~A.~Luty, A.~E.~Nelson and E.~Ponton,
  %``Gaugino mediated supersymmetry breaking,''
  JHEP {\bf 0001}, 003 (2000)
  [hep-ph/9911323].
  %%CITATION = HEP-PH/9911323;%%
  
  %\cite{Moroi:2012kg}
\bibitem{Moroi:2012kg} 
  T.~Moroi, T.~T.~Yanagida and N.~Yokozaki,
  %``Enhanced Higgs Mass in a Gaugino Mediation Model without the Polonyi Problem,''
  Phys.\ Lett.\ B {\bf 719}, 148 (2013)
  [arXiv:1211.4676 [hep-ph]];
  %%CITATION = ARXIV:1211.4676;%%
  %\cite{Yanagida:2013ah}
%\bibitem{Yanagida:2013ah} 
  T.~T.~Yanagida and N.~Yokozaki,
  %``Focus Point in Gaugino Mediation ~ Reconsideration of the Fine-tuning Problem ~,''
  Phys.\ Lett.\ B {\bf 722}, 355 (2013)
  [arXiv:1301.1137 [hep-ph]].
  %%CITATION = ARXIV:1301.1137;%%
  %7 citations counted in INSPIRE as of 14 Nov 2013
  
%\cite{Kawasaki:2008sn}
\bibitem{Kawasaki:2008sn} 
  M.~Kawasaki, K.~Nakayama, T.~Sekiguchi, T.~Suyama and F.~Takahashi,
  %``Non-Gaussianity from isocurvature perturbations,''
  JCAP {\bf 0811}, 019 (2008)
  [arXiv:0808.0009 [astro-ph]].
  %%CITATION = ARXIV:0808.0009;%%
  
  %\cite{Hikage:2012be}
\bibitem{Hikage:2012be} 
  C.~Hikage, M.~Kawasaki, T.~Sekiguchi and T.~Takahashi,
  %``CMB constraint on non-Gaussianity in isocurvature perturbations,''
  JCAP {\bf 1307}, 007 (2013)
  [arXiv:1211.1095, arXiv:1211.1095 [astro-ph.CO]].
  %%CITATION = ARXIV:1211.1095,;%%

%\cite{Dvali:1995ce}
\bibitem{Dvali:1995ce} 
  G.~R.~Dvali,
  %``Removing the cosmological bound on the axion scale,''
  hep-ph/9505253.
  %%CITATION = HEP-PH/9505253;%%
  %17 citations counted in INSPIRE as of 06 Nov 2013
  
  %\cite{Jeong:2013xta}
\bibitem{Jeong:2013xta} 
  K.~S.~Jeong and F.~Takahashi,
  %``Suppressing Isocurvature Perturbations of QCD Axion Dark Matter,''
  Phys.\ Lett.\ B {\bf 727}, 448 (2013)
  [arXiv:1304.8131 [hep-ph]].
  %%CITATION = ARXIV:1304.8131;%%
  %3 citations counted in INSPIRE as of 06 Nov 2013
  



%%%%%%%%%%%%%%%%%%%%%%%%%%%%%%%%%%%%%
\end{thebibliography}
\end{document}